\documentclass[preprint]{acmart}
\AtBeginDocument{%
  \providecommand\BibTeX{{%
    \normalfont B\kern-0.5em{\scshape i\kern-0.25em b}\kern-0.8em\TeX}}}

\copyrightyear{2023} 
\acmYear{2023} 
\setcopyright{rightsretained} 
\acmConference[FAccT '23]{2023 ACM Conference on Fairness, Accountability, and Transparency}{June 12--15, 2023}{Chicago, IL, USA}
\acmBooktitle{2023 ACM Conference on Fairness, Accountability, and Transparency (FAccT '23), June 12--15, 2023, Chicago, IL, USA}
\acmDOI{10.1145/3593013.3594133}
\acmISBN{979-8-4007-0192-4/23/06}
\begin{document}

\title{Reconciling Governmental Use of Online Targeting With Democracy}

\author{Katja Andrić}
\affiliation{
  \institution{The University of Edinburgh}
  \city{Edinburgh}
  \country{United Kingdom}}
\email{katy.andric@gmail.com}

\author{Atoosa Kasirzadeh}
\affiliation{
  \institution{The University of Edinburgh \& The Alan Turing Institute}
  \city{Edinburgh}
  \country{United Kingdom}}
\email{atoosa.kasirzadeh@ed.ac.uk}

\renewcommand{\shortauthors}{Andrić and Kasirzadeh}

\begin{abstract}
The societal and epistemological implications of online targeted advertising have been scrutinized by AI ethicists, legal scholars, and policymakers alike. However, the government's use of online targeting and its consequential socio-political ramifications remain under-explored from a critical socio-technical standpoint. This paper investigates the socio-political implications of governmental online targeting, using a case study of the UK government's application of such techniques for public policy objectives. We argue that this practice undermines democratic ideals, as it engenders three primary concerns --- Transparency, Privacy, and Equality --- that clash with fundamental democratic doctrines and values. To address these concerns, the paper introduces a preliminary blueprint for an AI governance framework that harmonizes governmental use of online targeting with certain democratic principles. Furthermore, we advocate for the creation of an independent, non-governmental regulatory body responsible for overseeing the process and monitoring the government's use of online targeting, a critical measure for preserving democratic values.\end{abstract}





\maketitle

\section{Introduction}

Online targeting, a fundamental aspect of the modern digital economy, involves customizing online products and services based on users' psychological profiles. These profiles are derived from algorithmic analysis of personal data, primarily acquired through the online monitoring and processing of data \cite{Beer2019,Koene2015}. Targeted ads, which have become an omnipresent feature of our online experiences, extend beyond merely displaying personalized ads for commercial items we have recently searched. As the Cambridge Analytica scandal revealed, online targeting has also permeated the political sphere, raising concerns about its broader implications \cite{schneble2018}.

In this paper, we explore the socio-political implications of the government's use of online targeting by examining a case study conducted by Collier et al. \cite{Collier2022}. This study uncovers that, since at least 2015, the UK government and law enforcement agencies have been employing targeted online advertising in various campaigns. Examples of such campaigns include a fire safety campaign ran by the Home Office that utilised Amazon data and targeted particular citizens through their Alexa speakers, or a National Crime Agency campaign, targeting young video gamers in their online environments with a goal of decreasing online crime activity. These campaigns aim to influence citizens' behavior by tailoring personalized online advertisements according to their psychological profiles. Prior to Collier et al.'s research \cite{Collier2022}, civil society in the UK was largely unaware of the prevalence and scope of such practices.

Apart from the work by Collier et al. \cite{Collier2022}, there is a notable gap in research regarding the socio-political implications of governments' use of online targeting.\footnote{Some discussions of similar practices can be found within the field of strategic communication. However, much of how governments employ online targeting remains implicit. For example, see \cite{michelsen2019} for a good overview of the field of strategic communication in the era of Big Data and \cite{hyland2021} for a discussion of governmental communication during the COVID-19 pandemic, which included some online targeting.} Given the unique relationship between governments' use of online targeting, legitimate power, and the vast number of individuals affected by this contentious practice, there is an urgent need for interdisciplinary research to address the socio-political implications of governmental use of online targeting.

In this paper, we present a critical and philosophical analysis of the socio-political challenges posed by the UK government's use of online targeted advertising to achieve public policy objectives. We investigate the contested relationship between this practice and the fundamental principles of democracy. We explore potential strategies to reconcile the government's use of online targeting with the core values and tenets of a democratic society. We emphasize that this issue warrants significant attention in terms of designing appropriate regulatory and governance mechanisms, particularly as generative artificial intelligence (e.g., large language models) has the potential to significantly enhance online targeting by accelerating the process, reducing costs, and improving the quality of content production \cite{weidinger2022taxonomy}. This underscores the need for a timely and thorough examination of the implications of such governmental use of online targeting within the context of democracy and public policy.

Thus far, the majority of philosophical literature on online targeting within the political domain has primarily concentrated on its influence on personal autonomy during voting and the negative repercussions for democracy \cite{Susserand2019, Susser20192, zuiderveen2018, barnhill}. Nevertheless, several other essential democratic values and principles, including equality, transparency, and privacy, frequently remain underexplored in this context. This paper endeavors to examine and highlight the often-neglected impacts of the governmental use of online targeting on these values and principles, thereby offering a more comprehensive perspective on the subject. The paper will proceed as follows.

In Section 2, we define online targeting and briefly delve into its history and applications. In Section 3, we present Collier et al.'s study \cite{Collier2022} of the UK government's use of online targeting, analyzing various past campaigns undertaken by the government. In Section 4, we highlight three key problems posed by governments utilizing online targeting: the Transparency Problem, the Privacy Problem, and the Equality Problem. We will demonstrate that each of these concerns conflicts with at least one fundamental aspect of democracy, as outlined by Lever \cite{lever2006}: (i) enabling citizens to be informed participants in law-making and electing representatives; (ii) upholding citizens' civil, socioeconomic, and political rights; and (iii) safeguarding citizens' equality and freedom. In Section 5, we argue that online targeting by governments can still be reconciled with democratic principles and values. Drawing on Züger and Asghari's AI governance framework \cite{zuger2022}, we advocate for the establishment of an independent institution to oversee campaign design and deployment, serving as a check on government power. We show that this solution can, in principle, effectively address the three problems raised in Section 4. The final section concludes the paper.

\section{What is Online Targeting?}

In today's digital world, an immense volume of personal data is accessible, encompassing not only basic details like names, email addresses, and birth dates but also extending to information derived from social media activity, sexual orientation, health records, search history, and purchasing habits. This data is collected, generated, stored, and managed by various entities, including commercial companies like Google, Facebook, and Amazon, as well as data brokers who specialize in aggregating and selling data to third parties \cite{anthes2015,Crain2018}.\footnote{Data brokers often operate behind the scenes, amassing information from numerous sources to create comprehensive psychological profiles of individuals, which are then traded within the data market for various purposes.}

Psychological profiles in online targeting refer to the comprehensive digital representation of an individual's personality, preferences, and behaviors, derived from their online activities \cite{boerman2017online}. These profiles are generated using data mining and machine learning algorithms that analyze various data points, such as browsing history, social media interactions, and online purchases, to infer users' interests, habits, and tendencies. The profile-building process often involves combining data from multiple sources, employing machine learning techniques to identify patterns, and categorizing individuals based on shared characteristics. These profiles are designed to forecast users' behavior and decision-making processes in various situations \cite{Koene2015}.

By constructing such detailed portraits of users, advertisers and commercial companies can tailor their content, messages, or campaigns to resonate with specific target audiences, ultimately enhancing the effectiveness of their online engagement \cite{Beer2019} and advancing their business goals \cite{Susserand2019,matz2017psychological}. This process of adapting online content to align with the psychological profiles of users is known as \textit{online targeting}.

While online targeting is a relatively recent development, its offline counterpart has a more extensive history. The practice of influencing individual or group behavior by appealing to their psychology predates the advent of the internet or algorithm-driven profiling. A notable example, as described by Halpern \cite{halpern2015}, occurred in the $18^{th}$ century when Frederick the Great successfully promoted the consumption of potatoes --- a formerly unpopular and bland vegetable --- in order to stave off famine. He accomplished this by establishing a guard around the royal potato fields and publicly expressing his admiration for the crop. This strategy piqued public interest and facilitated the widespread popularity of potatoes throughout Prussia. Later, in the $20^{th}$ century, Edward Bernays leveraged psychological principles to influence public opinion and behavior, making him a key historical figure in the practice of manipulating individual or group behavior by appealing to their psychology \cite{bernays1947engineering,bernays1928manipulating}. Today, such behavior can be examined through the lens of social and cognitive psychology, disciplines that were first combined with economics in the early 20th century to give rise to the field of behavioral economics.

Behavioral economics investigates the consequences of human cognitive limitations on decision-making within markets \cite{mullainathan2001}. Classical economics traditionally portrays individuals as rational decision-makers who consistently select the optimal choice after conducting thorough cost-benefit analyses, uninfluenced by extrinsic factors or emotions. In contrast, insights from various disciplines have revealed the impact of cognitive biases, emotions, perceptions, heuristics, and social contexts on rationality \cite{halpern2015}. Consequently, economists have come to acknowledge that the way the choice options are presented can significantly influence the choices we make.

In 2008, renowned scholars Richard H. Thaler and Cass R. Sunstein, the former having been awarded a Nobel Prize for his significant contributions to behavioral economics, introduced the concept of \emph{choice architecture} \cite{thaler2008}. This idea pertains to the environments in which we make decisions, encompassing the range of options available to us, the manner in which they are presented, and the entities presenting them. Thaler and Sunstein maintain that these choice architectures inherently influence the decisions we make. As a result, if we aim to guide someone towards a specific choice, we can modify their choice architecture through the application of \emph{nudges}. As defined by \citet[p.6]{thaler2008}, a nudge is an aspect of the choice architecture that predictably alters people's behavior without forbidding any options or significantly changing their economic incentives. This concept suggests that subtle changes in choice architecture can have a significant impact on decision-making. A prime example of this would be strategically placing healthy food options at eye level in a grocery store, while positioning unhealthy alternatives out of sight, in order to boost sales of healthier choices.

Online targeting operates on a principle similar to that of its non-digital counterparts, such as grocery stores, wherein choice structures are manipulated to guide user behavior towards a desired outcome. What sets online targeting apart, however, is the capacity to accurately segment audiences based on their algorithmically generated profiles. Furthermore, it allows for dynamic adjustments to choice structures tailored to each individual user, taking into account their personal preferences \cite{susser2019}. User-facing platforms, which serve as primary channels for online targeting, are meticulously optimized through extensive user experience (UX) testing methods and the application of behavioral insights. This optimization aims to maximize user attention, resulting in heightened engagement with targeted content and the conversion of this attention into revenue or other desired behaviors \cite{bakir}. This characteristic sets online targeting apart from offline targeting, as it enables the deliberate and strategic delivery of customized nudges to each user, informed by an understanding of the nudge's impact on that specific individual. As a result, these personalized nudges prove more potent, allowing those employing targeting tactics to accomplish their business objectives with greater efficiency and effectiveness.

The pervasive use of online targeting, especially in targeted advertising --- which constitutes 79\% of all online advertising \cite{miller2018} --- is hardly surprising in today's digital landscape. Internet companies frequently employ Chief Behavioral Officers and Choice Architecture Engineers to shape their customers' behavior \cite{Kirchhof2021}. As a foundational component of the modern internet economy \cite{Beer2019}, online targeting contributes to enhanced sales of goods and services. Additionally, it enables numerous apps and websites, ranging from Facebook to The Guardian, to provide low-cost or free services by relying on revenue generated through targeted ads \cite{johnson2022}.

Online targeting techniques, which have proven effective in shaping behavior, are not only confined to the commercial realm. These methods have infiltrated the political sphere, where they have been employed in political campaigns to sway voters' decisions \cite{crain2019, dommett}. The infrastructure of online targeting has thereby enriched the longstanding practice of political marketing, which dates back to the 1980s \cite{wring}. The tactics of online political targeting were reported during Barack Obama's groundbreaking 2008 presidential campaign which involved identifying supporters and persuading undecided voters \cite{plouffe2010,farman}, but gained even greater notoriety during Donald Trump's 2016 presidential campaign \cite{Bomelburg2021, bakir}. In the latter campaign, the Trump team enlisted the expertise of Cambridge Analytica, a British political consultancy firm. This company harvested data from an astonishing 87 million Facebook users through a seemingly innocuous personality quiz \cite{schneble2018}. With this wealth of information, Cambridge Analytica crafted intricate psychological profiles of users and deployed emotionally manipulative, personalized advertisements to influence their voting behavior in favor of Trump \cite{Bomelburg2021, bakir}. This striking example underscores the pervasive reach of online targeting techniques. It also highlights their potentially profound impact on political outcomes.

Online political targeting is not limited to the United States, as political parties across Europe have also utilized these techniques \cite{zuiderveen2018}. The UK's House of Commons Select Committee on Digital, Culture, Media and Sport has described online political targeting as a "democratic crisis" \cite[p.51]{dsmc2018}. As we intend to show in this paper, this crisis has deepened, with the government now using the online targeting infrastructure for public policy purposes, which can undermine fundamental principles of democracy if used inappropriately. In the following section, we examine a case study that demonstrates how the UK government's utilization of online targeting can potentially compromise certain democratic principles.

\section{Governmental Use of Online Targeting}

A recent study by Collier et al. \cite{Collier2022} has uncovered a striking finding - the UK government and law enforcement agencies are utilizing online targeted advertising to influence citizens' behavior and achieve public policy objectives. This use of online targeting presents unique socio-political implications that warrant further exploration. To understand the potential consequences, it is essential to first grasp the nature of the UK government's employment of online targeted advertising.

The UK government has a history of attempting to change the behavior of its citizens in order to achieve public policy goals. Following the establishment of the Behavioural Insights Team (also known as \textit{Nudge Unit}) in 2010, the government has been openly applying nudge theory to public policy \cite{Collier2022}. The Team's objective is to influence people's choices by designing appropriate incentives or obstacles, thereby encouraging the desired options, all while incorporating insights from behavioral economics \cite{Halpern2017}. The Nudge Unit's inception was fueled by the growing popularity of behavioral change techniques combined with social marketing practices among UK public policy makers, who had been experimenting with the application of commercial marketing principles to promote public goods since 2004 \cite{pykett}. Examples of the Team's initiatives include adjusting tobacco prices to deter individuals from purchasing it or incorporating carefully crafted tax prompts in letters to taxpayers to encourage prompt payments (e.g., "most people pay their tax on time") \cite{Halpern2017}.

While influencing citizens' behaviour for the purpose of achieving public policy objectives is not a new practice for the UK government, what \textit{is} a new practice is combining these behaviour change strategies with the infrastructures of online targeted advertising.\footnote{Adapting commercial marketing technologies, such as online targeted advertising, to programs that are designed to influence the behaviour of targeted individuals for their benefit as well as wider social benefit, is what Andreasen \cite{andreasen} terms \textit{social marketing}.} As explained in Section 2, those infrastructures involve data gathering and processing by machine learning algorithms to create profiles of individuals and groups so that messages specifically tailored to each profile can be created and delivered to those deemed susceptible. Collier et al.'s research \cite{Collier2022} shows that government departments are combining operational data gathered and produced by state institutions and the associated systems of classification and profiling of social groups (e.g. the needs, risks and vulnerabilities of groups such as patients, immigrants and welfare recipients) with data gathered and produced by commercial companies and the internet economy (e.g. clicks, page visits, shopping habits, social media activity, and online social interactions). The hybridisation of the two categories of data and their algorithmic processing allows for both a wide and a deep insight into UK's communities and individuals.

During the past decade, the UK government ran numerous campaigns at the heart of which was online targeting. Collier et al. \cite{Collier2022} map those into three distinct modes of operation, which we term (i) the \textit{minimally targeted mode}, (ii) the \textit{maximally targeted mode}, and (iii) the \textit{outsourced mode} for ease of reference. The first two modes are delivered by government and law enforcement agencies and differ in the level of sophistication of the targeting used. In contrast, the last mode involves outsourcing the services of private sector companies but is on par with the maximally targeted mode in terms of targeting sophistication.

The first and the least sophisticated mode, i.e., the \textit{minimally targeted mode}, amounts to simply extending the advertising scope to online spaces through online advertisement buys. As the name suggests, this mode is minimally targeted as it does not involve much audience segregation or reliance on individuals' profiles but is targeted at entire population groups. It involves actions such as running advertisements on Tiktok to reach younger audiences.

The second and significantly more sophisticated mode of operation, which we term the \textit{maximally targeted mode}, leverages targeting to reach specific groups or individuals and informs the design of the run adverts. This mode employs algorithms that enhance personalization and relies on a network of government entities led by the Government Communication Service to develop nationwide behavior change strategies. In addition to implementing multi-site and single-site campaigns to change behavior through tailored messages, this mode also includes countering misinformation online and protecting the government's reputation from negative messages. One notable example cited by \citet[p.7]{Collier2022}. is a fire safety campaign by the Home Office, where the department utilized purchasing data from individuals who recently bought candles on Amazon and targeted them with fire safety messages through their smart speakers, such as Alexa.

The maximally targeted mode also involves leveraging the maximally targeted technologies and methods employed by law enforcement for preventive purposes. A prime example of this is the CYBER CHOICES preventative diversion program ran by the UK National Crime Agency (NCA) in collaboration with behavioral psychologists \cite{Collier2022}. This initiative utilized Google and YouTube advertisements targeted towards UK adolescents between the ages of 14 and 20 who were identified through NCA surveillance as potentially interested in gaming. The ads would appear whenever they searched for cybercrime services and warn them of the illegal nature of purchasing such services and the consequences they could face if they did so. As a part of this campaign, NCA officers also visited the identified "targets" at their homes, discussed their online behaviour with their parents, and invited them to workshops organised and ran by NCA. The goal of those workshops was twofold. Firstly, the individuals were taught the skills required to turn their illegitimate interests into a legitimate career. Secondly, NCA used the workshops to gather data to optimise the design of further targeted ads. There is evidence that this particular campaign was successful in reducing the rate of cybercrime \cite{collier2021}.

The third and final mode of operation, the \textit{outsourced mode}, entails entrusting the entire process of designing, developing, and executing campaigns to private sector companies \cite{Collier2022}. One instance is SuperSisters, a Muslim online lifestyle platform established by J-Go Media in 2015, aimed at young British Muslim girls. Although marketed as a platform for sharing and creation of empowering content, the project sparked controversy when it was revealed that it was covertly funded by a government counter-extremism arm and that the content on the website was carefully curated to counteract what the state deemed to be "overtly Islamic" \cite{iqbal2019}. Another example includes a UK Home Office-supported knife prevention campaign, targeting young Black individuals residing in London's deprived neighborhoods, created by FCB Inferno and All City Media \cite{Collier2022}. The campaign's offline component included messages displayed on takeout boxes in fried chicken restaurants, based on police data which indicated that Black individuals commit more knife crimes and perpetuated a racist stereotype that they consume fried chicken \cite{webster2019}. The online component aimed at young Black males living in impoverished areas of London drew upon the same data and stereotype.

\section{Socio-political Issues and Online Targeting's Contested Relationship with Democracy }

Some of the examples outlined in the previous section may cause discomfort. In this section, we will pinpoint some of the social and political issues that arise from the use of online targeted advertising by democratic governments to achieve public policy goals. Clearly formulating these issues will not only concretize the discomfort, but it will also lay the groundwork for a deliberation on the legitimacy of such practices and, if necessary, their proper form. The examination of these questions will be the focal point of the upcoming section. For the moment, our attention will be directed toward highlighting the challenges associated with this procedure.

In the context of governmental use of online targeted advertising, the primary socio-political concern is the potential for abuse of power and erosion of democratic values.\footnote{Additional ethical issues and potential solutions surrounding the practice of targeted advertising have been extensively explored in literature, including works by Thaler and Sunstein \cite{thaler2008}, Hansen and Jespersen \cite{hansen2013}, Wilkinson \cite{wilkinson2013}, and Nys and Engelen \cite{nys2017}. It is important to note that some of these issues are not exclusive to online targeting or governmental use of it, as non-digital nudging also raises similar concerns. For the lack of space, we will not be covering the non-digital instances in this paper.} According to Susser \cite{susser2019}, autonomy refers to a person's ability to make decisions based on their own personal values and beliefs. Online targeted advertising undermines autonomy by intentionally and covertly manipulating individuals through exploiting their decision-making vulnerabilities and cognitive biases \cite[p.4]{Susserand2019}. This manipulation often goes unnoticed, as individuals are unaware of the influence on their decision-making. In line with this concern and as a result of governmental online targeting, individuals are left open to being molded into the ideal citizens according to the government's preferences. While we acknowledge the negative impacts on autonomy through manipulation, governmental use of online targeting raises additional and novel socio-political issues that have not yet been thoroughly analyzed. This paper will focus specifically on such socio-political concerns.

At the core of the socio-political concerns specific to governmental online targeting is the practice's contested relationship with democracy and democratic values. So far, philosophical literature tackling the relationship between online targeting and democracy has focused almost solely on the detrimental effect online targeting has on one's autonomy during voting, the manipulative nature of this practice and the dangers it poses for democratic elections \cite{Susserand2019, Susser20192, zuiderveen2018, barnhill}. However, governmental use of online targeting for public policy goals brought to light by Collier et al.'s research \cite{Collier2022} expands the known scope of online targeting in the political domain beyond using targeted adverts on citizens during election campaigns. This novel use of online targeted advertising undermines other, overlooked, features central to democracy besides voting. These are the ones we will tackle.

According to Lever \cite{lever2006}, democracy comprises three key elements: (1) allowing citizens to be informed participants in decision-making processes, including the creation of laws and the election of representatives; (2) guaranteeing and preserving civil, socioeconomic, and political rights; and (3) ensuring equality and freedom for all citizens. However, the UK government's practice of using tailored online advertisements, particularly through maximally targeted and outsourced methods, violates --- to a certain degree --- all three of these democratic principles. We argue that this practice raises three major issues - the Transparency Problem, the Privacy Problem, and the Equality Problem - that directly challenge one or more of the key elements of democracy.

\subsection{The Transparency Problem }

Transparency is generally understood to be one of the central principles of democracy \cite{rosendorff2006}. If citizens do not have the ability to freely access information, they cannot keep the government accountable for their actions and decisions. Moreover, they cannot make informed decisions at the ballot box or actively participate in other democratic processes, such as publicly questioning and criticising the government's decisions. The lack of transparency about the governmental use of online targeting undermines the feature (1) of democracy mentioned above: enabling its citizens to participate in the determination of laws and the election of their representatives. In this subsection, we will show that the UK government's use of online targeting suffers from a lack of transparency. We call this the Transparency Problem.

The Transparency problem can be viewed from different angels. Firstly, there is a general lack of governmental transparency about the practice of online targeted advertising. The government has never published a comprehensive list of online targeting campaigns it has created itself or outsourced from the private sector. While some scarce information is available on different governmental bodies' websites, most of the information about the campaigns comes from the documentation submitted to various industry awards by third-party agencies hired by the government to develop and run the campaigns.\footnote{We confirmed this through personal communication with Ben Collier, 2022.}. From government records only, it is uncertain how many campaigns the government has run so far, who the campaigns targeted, or what the content of those campaigns was.

Apart from the former lack of transparency concerning the practice, online targeted advertising suffers from an inherent lack of widespread transparency. As Collier et al. \cite{Collier2022} argue, targeted adverts are normally only seen and intended to be seen by those targeted, meaning that most of the population will never encounter them. In contrast, non-targeted governmental campaigns delivered either online or offline are, in principle, visible to the entire population (including the press) who can scrutinize and challenge them. The lack of extensive visibility of targeted adverts reduces the capability for broader scrutiny. Consequently, it reduces the public's ability to hold the government accountable for its actions.

Moreover, determining the effectiveness of targeted governmental advertisements is challenging since it cannot be measured by directly observable outcomes, such as sales conversion, as is the case with commercial targeting \cite{Collier2022}.  This not only makes it difficult to justify the measure, but also reduces the citizens' ability to hold the government accountable, as they must not only know what the government is doing, but also whether it is achieving satisfactory results.\footnote{The field of public relations also faces the challenge of determining the success of its campaigns due to the lack of readily quantifiable metrics for success. Some efforts have been made to establish a framework for tracking and measuring the impact of public relations campaigns in the field, which could be relevant to governmental online targeting. For example, see \cite{bakir2}, \cite{Michaelson} and \cite{plowman}}

A final blow to transparency is delivered by the fact that machine learning algorithms underpinning the structure of targeted advertising suffer from an inherent opacity problem. This means that it is not always possible to know why and how --- in non-mathematical terms --- an algorithm reached a specific prediction \cite{burrell2016machine,kasirzadeh2021reasons,kasirzadeh2021use,gunther2022algorithmic}. The implication is that citizens are often unable to obtain an explanation for why they were targeted with a specific advert. This issue is exacerbated by Collier's assertion that, even upon request, the government will not disclose targeting data, which is the data used to identify individuals for a particular advert.\footnote{This was revealed to us in a personal conversation with Ben Collier in August 2022.} Without sufficient information being provided to the relevant stakeholders about the algorithmic systems that make decisions about them, it is challenging to envision any meaningful discussion of the ethical concerns raised by the behavior of the system \cite{Diakopoulos2020}.

\subsection{The Privacy Problem}

The concept of privacy carries various interpretations \cite{dwork2006differential,finn2013seven,moore2008defining,liu2021}. A significant definition posits privacy as the capacity to control our personal information. This control enables us to manage others' knowledge about us, thereby establishing varying degrees of intimacy with different individuals or groups \cite{solove2008}. Crucially, the freedom to disclose our personal details at our discretion and to chosen recipients safeguards our autonomy and dignity. In instances where our personal data is collected and analyzed, the exercise of informed consent ensures we retain control. Conversely, data collection and analysis conducted without our informed consent infringes upon our right to privacy.

The right to privacy is a fundamental human right recognised by democratic countries \cite{thomson1975right,cohen2013privacy}. To illustrate, in the UK, this right is protected by the \textit{\citeauthor{humanrights}}. This right offers protection to citizens from undue and illegal governmental surveillance. Moreover, it fosters an environment where individuals feel secure to explore, express their beliefs, and cultivate their interests. In this subsection, we will make the argument that the UK government has, in some instances, collected data on its citizens without obtaining informed consent. This action can be seen as an infringement on the citizens' right to privacy, a concern we refer to as the Privacy Problem. This problem not only conflicts with Lever's second principle of democracy - the preservation of citizens' civil, socioeconomic, and political rights - but also compromises the citizens' capacity to make genuinely independent political decisions, contradicting the first principle. Let us unpack this problem.

Privacy and consent are closely interrelated. The act of consenting to data collection and analysis allows you to exert control over your personal data. Consent, viewed as a normative concept, can make an act that would otherwise be impermissible, permissible by facilitating the transfer of rights and obligations between involved parties \cite{letajones2020}. For consent to be morally transformative, it must be informed. In the context of data, this necessitates the provision of the following: (i) an explicit description of the potential uses and restrictions of your data, (ii) a specific definition of the scope within which your data can be utilized, (iii) ample information for the consenting party to comprehend what they are agreeing to, (iv) a range of free-to-choose options, and (v) a mutually equitable treatment between both parties \cite{letajones2020}. In the UK, the personal information of individuals is safeguarded by the \textit{\citeauthor{dataprotection}}, which outlines six principles of data protection that all parties responsible for using personal data must adhere to. For example, the first data protection principle requires that consent is obtained from the individual for the information to be collected and processed.\footnote{Notably, the Act lists certain exceptions, but a full interpretation of these is beyond the scope of this paper.}

If a party plans to use your data in a way that was not disclosed during the initial consent process, this original consent becomes invalid and a fresh consent is required. For example, if a party intends to use your postal code for a different purpose than the one you initially agreed to, such as determining insurance rates, they must seek your permission again \cite{Andreotta2022}. Similarly, if a party intends to combine your data with another dataset (for both of which you have given individual consents), they must ask for your renewed consent. This is because the merger could generate new information that was not anticipated when you initially gave consent.

The UK government's use of targeted advertising raises concerns regarding the validity of consent due to not disclosing relevant information in both of above senses. While the government has sought permission from citizens to gather administrative data, it neglected to disclose that this data could be utilized for profiling and online targeted advertising. It is probable that the surveys used to collect such data contained a clause explicitly stating the data would not be used for marketing purposes. As a result, the consent previously acquired is flawed and fresh consent must be obtained. Additionally, the lack of transparency surrounding this practice leaves many unaware that their online activity, such as Facebook likes and Amazon purchases, could be leveraged to fulfill public policy objectives. This data, when merged with census data, generates new information for which proper consent was not sought. Consequently, this practice seems to be illegitimate.

The Privacy Problem, besides raising questions about consent validity, also impedes UK citizens' full participation in democratic processes. Here is why. The erosion of privacy harms not only those whose privacy is at stake, but also democracy itself \cite{lever2006,debrabander_2020}. Privacy creates a safe space for individuals, facilitating the growth of their opinions, the exploration and pursuit of diverse interests, and the freedom to make decisions without fear of judgment or backlash. Being constantly monitored, with the potential of being categorized into risk groups by the government or targeted with intimidating adverts, and in extreme cases, even receiving home visits for merely searching certain terms (as seen in the CYBER CHOICES campaign), can instigate self-censorship and make people feel unsafe in their own country. This is especially true for historically disadvantaged groups, who may feel particularly unsafe and skeptical towards the government \cite{lever2006}.

Governments have long engaged specific subsets of the population through various forms of offline advertising and public policy campaigns. For instance, public health campaigns have often targeted smokers, warning them of smoking dangers via pamphlets distributed in healthcare facilities and television commercials. However, there is a distinct contrast between such campaigns and those founded on the framework of online targeting. In the case of offline campaigns, individuals are aware they are the target because they identify with the group in question. On the other hand, when it comes to online governmental targeting, individuals receive specific advertisements because they have been personally identified by the government as relevant recipients. This identification process is made possible by continuous governmental monitoring of the individual's online activities, to which they likely did not consent. Furthermore, this kind of targeting grants governments access to previously inaccessible areas --- private homes. For example, whereas in the past, you might have encountered a fire safety poster while commuting to work, now, you might be greeted by a message about fire safety from Alexa when you return home because you purchased a candle on Amazon the previous week.

Online targeting by government agencies is significantly more invasive than traditional methods, largely due to the opaque nature of the data collection process. This lack of transparency can exacerbate feelings of paranoia and vulnerability. The core of the chilling effect we are discussing lies in this disparity. Persistent surveillance fosters an environment of constraint, where individual autonomy is violated \cite{benn_lazar_2022}. For a democracy to thrive, it necessitates independent and autonomous decision-makers, a condition which becomes challenging to fulfill in the absence of privacy. Consequently, privacy appears to be a crucial component in exercising our democratic rights, including political choice \cite{lever2006}.


\subsection{The Equality Problem}

The third socio-political concern central to our discussion is what we call the Equality Problem. This problem relates to the possibility that government-led targeted advertising may undermine the principles of equality and justice upheld by democracy, thereby violating the third democratic feature outlined by Lever.

In an effort to identify the target audience for a campaign, the UK government and law enforcement construct a profile of the ideal target and advertise to individuals who fit this profile. This profile is created using available data and is generated algorithmically. However, it has become increasingly clear that AI algorithms have the potential to perpetuate existing wrongful social inequalities \cite{buolamwini2018gender,eubanks2018automating,obermeyer2019dissecting}. This is largely due to the data used to develop these algorithms often reflecting the biases and disparities inherent in society. The algorithms look for patterns in the data, but if the data reflects wrongful inequalities, such as over-policing of Black neighborhoods, then the output of these algorithms risks perpetuating existing social hierarchies.

Consequently, this approach has the potential to exacerbate the marginalization of already disadvantaged groups. It could erroneously place these individuals into risk categories, not because of their actions, but due to systemic discrimination. This sort of bias, targeting protected characteristics like race, sex, gender, and disability, is not only unethical but also unlawful according to the authors of \textit{\citeauthor{equalityact}}. A notable instance of this is the targeting of young Black males residing in economically disadvantaged areas of London in a knife crime prevention campaign, a topic elaborated upon in Section 3.


Algorithmic bias can lead to discrimination, but it's not the only source. As AI algorithms unearth patterns within data sets, novel forms of discrimination can arise, not necessarily linked to traditionally protected attributes \cite{wachter2021}. People can be placed into "ad hoc" categories, like dog owners or video gamers, and subsequently face unfair treatment compared to those outside these groups. This is because the algorithm had detected a correlation between owning a dog or playing video games and certain behaviors. However, owning a dog or playing video games may not be a normatively acceptable basis for forming a group around them if the formation occurred as a result of a spurious correlation found in the data or if the statistical correlation is insufficiently significant \cite{wachter2021}.

The use of online targeting by the government is not only leading to unfair and discriminatory outcomes for individuals, but it also poses a deeper problem with regards to equality. It creates a shift in power dynamics within society, giving the government more control over its citizens than is desirable in a democratic society \cite{zuboff2015big}.

The connection between knowledge and power cannot be ignored. The government's collection and analysis of individuals' data gives them a greater ability to influence and control their citizens \cite{veliz2020}. This raises concerns about potential misuse of such power, as warned by Königs \cite{konigs2022}. Furthermore, it raises questions about who has the authority to shape society and determine its priorities \cite{Collier2022}.

The question therefore persists: Can the practice of targeted governmental advertising ever align with democratic principles and values? The following section outlines steps towards answering this inquiry.

\section{Reconciling Online Targeting with Democracy }

In light of our previous discussions, it is clear that online targeting can be an effective strategy for achieving certain goals. This approach allows the government to identify and engage with specific subgroups more effectively, offer tailored resources to address public issues, and optimise the use of their limited resources. However, its employment by the UK public sector has provoked serious questions about transparency, privacy, and equality. In order to align its use with democratic principles, the government must address these concerns. While we acknowledge there is no easy fix, we propose a few steps towards a potential solution.

First, we suggest requirements for the design and execution of online governmental targeting campaigns that are in alignment with democratic values. We will reference the recent AI governance framework developed by Züger and Asghari \cite{zuger2022} as our initial guideline. Second, we propose the creation of an independent body to monitor the development and implementation of these campaigns. This institution will ensure compliance with the defined requirements and provide guidance to government officials.

The use of AI for social benefits has been gaining momentum in recent years. Numerous applications of AI are now directed towards addressing issues that impact human life and well-being \cite{tomavsev2020ai,cowls2021definition}. Considering that public policy should embody public interests \cite{dejersey2003}, utilizing AI-powered online targeting to achieve these policy goals is another way in which AI can serve the public interest and contribute to societal good. This usage of AI-powered online targeting to fulfill public policy objectives falls within the ambit of an AI governance framework that prioritizes public interest. According to Züger and Asghari \cite{zuger2022}, there are five prerequisites for a system to align with the public interest: (1) public justification, (2) focus on equality, (3) inclusion of a deliberation/co-design process, (4) implementation of technical safeguards, and (5) commitment to openness for validation.

To successfully address the three problems we highlighted in the previous section, all five requirements must be met. The first three requirements are context-sensitive and their fulfillment will differ depending on the project. Requirements 4 and 5, being technical supplements, are less dependent on the specific nature of the project.

\subsection{Public Justification}

Züger and Asghari \cite{zuger2022} propose that for an AI-based solution to be recognized as serving the public interest, it must possess a normative democratic justification that is widely accepted by the public. This justification should include a lucid explanation of the issue that the AI solution seeks to address and why it is superior to alternative solutions. This stipulation is anchored in the philosophical views of public interest presented by Habermas \cite{habermas2022neuer}, Held \cite{held1970}, and Bozeman \cite{bozeman2007}, who assert that the public should determine what is in their best interest on an individual case basis.

In terms of the UK government's approach to online targeting, it is crucial for the government to maintain transparency. This involves disclosing any online targeting campaigns to the public, clarifying why this particular method was chosen, and detailing the operational aspects of the campaign, such as the ad content and target demographic. By doing so, the government can counteract the prevalent lack of transparency in this practice. To further enhance accountability, the government should also consistently update and provide a predictive measure of the campaign's efficacy, as well as quantifying interactions with the ads. Nevertheless, while conducting these processes, it is of utmost importance to safeguard individual identities by anonymising the data.

Finally, the justification should also provide easily understandable information about opaque algorithms. By adequately fulfilling these requirements, the Transparency Problem can be avoided.

\subsection{Equality}

Züger and Asghari \cite{zuger2022} assert that serving equality (and at the very least not hurting it) must be the most important normative goal of a solution that aims to promote public interest. Thus, they argue that any public interest AI-based solution must find a way to solve the problem of algorithmic bias and not create unwanted power imbalances in society. The second requirement is also based on previous scholarship. More precisely, the work of a legal scholar Feintuck \cite{feintuck2004}, who argues that something can be in the public interest only if it promotes equality of citizenship.

To satisfy this requirement, the UK government should not run any campaigns that discriminate against its citizens (whether based on some protected characteristic stemming from algorithmic bias or on spurious correlations found in the data), reproduce harmful social hierarchies or create new power imbalances. By fulfilling this requirement, the government would solve the first part of the Equality Problem - discrimination by the government. There remains the second part of the problem - unwanted power asymmetry. The proposed preliminary solution for this issue will be presented at the end of this section, where we suggest the formation of an independent institution tasked with monitoring these practices.

\subsection{Deliberation/Co-Design Process}

Züger and Asghari \cite{zuger2022} argue that to determine what is in the public interest for a public at a given time, the public must be involved in the system's design through the process of deliberation. The process can take any form, from online documentation to interviews with citizens. Without public deliberation about the public's interests, the team of developers will have to assume the interests of others which can easily lead to harmful mischaracterisation, unintended consequences and public rejection of the project \cite{zuger2022}. Therefore, they argue that those who will be affected by the system must have their say. As with the previous two requirements, Züger and Asghari root this requirement in an existing philosophical theory. According to Bozeman \cite{bozeman2007}, who draws from Dewey \cite{Dewey2012}, the individuals who form the public can only determine what is in the public interest through public deliberation by expressing their personal views, listening empathetically to others and reaching a compromise which benefits everybody. 

Such a highly democratic approach is available to the UK government, although it is rarely followed since it requires significant resources. However, a 2021 outsourced campaign by Police Scotland called Breaking the Cycle of Fear, whose goal was to reduce violence in the most deprived areas of Glasgow and Dundee, showed why such a process is desirable, despite the required resources. During the design process, interviews were conducted with individuals from targeted areas who had managed to escape the cycle of violence. The interviews aimed to understand what life in such areas looks like, how one gets involved with violence, and how one escapes it. Data gathered through these interviews informed the design of a targeted advert - in this case, a short movie\footnote{The movie can be viewed here: \url{https://www.youtube.com/watch?v=APUfXvepLQQ\&t=3s}}- which was assessed for credibility and realism by the interviewees. 

In three months, almost 500 targeted people reached out to the Scottish Violence Reduction Unit asking for help in escaping the cycle of violence themselves. Moreover, the campaign received only positive criticism from the targeted population, who did not report feeling marginalised. Had the previously discussed knife crime reduction campaign followed a similar approach, it could have avoided relying on harmful stereotypes and further marginalising the communities they were trying to reach. Therefore, governments should follow this approach in future online targeting-based campaigns as this will significantly help solve the Equality Problem.

\subsection{Technical Safeguards and Openness to Validation}

Züger and Asghari \cite{zuger2022} argue that AI-based systems need to implement technical safeguards, including data quality and system accuracy, data privacy, and safety and security. That is, the data fed into the AI systems needs to be free of bias and of high quality so that the outcomes are accurate, can be validated and serve equality. Satisfying this technical safeguard would take us closer to solving the Equality Problem. Further, data protection and privacy laws must be complied with. This includes obtaining informed consent from everyone whose data is being used, which would solve the consent part of the Privacy Problem. Finally, it must be ensured that the system is secure and robust to eliminate malfunctions, unintended functionalities and security breaches.

The next requirement Züger and Asghari \cite{zuger2022} propose amounts to having the entire system, including the design process, open to validation of others. They note two main reasons for this requirement. Firstly, any system that impacts the public at large may cause unintended harm, regardless of the good intentions of its makers. Secondly, any system that claims to be in the public interest should follow the basic democratic norm of transparency, allowing those impacted by the system to review all decisions made by the systems' makers and the workings of the technology to ensure that its mechanisms are democratic \cite{zuger2022}. This would help with the explainability of the system and, consequently, people's trust in the system. Thus, if the UK government satisfied this requirement, it would take another step toward solving the Transparency Problem.

\subsection{Independent Institution}

In theory, satisfying these five requirements could be left entirely to the government's discretion while trusting that they will behave ethically. However, there would be no way of ensuring that the government is sincere in following the guidelines without an independent organization overseeing the entire process and serving as a check on the government's power. Moreover, the government officials who currently run these campaigns do not always possess the necessary knowledge to understand the harms a campaign may cause, the technical workings of the algorithms behind the campaigns or the regulation that needs to be followed. Interviews with UK public officials reveal that online targeting campaigns, especially at a local level, are often designed without much consideration of the ethical issues they may raise, prior research or planning. For example, Collier and Wilson \cite{collierwilson} report an insider to a UK government-led counter-radicalisation campaign describe their approach as "throwing things at a wall to see what sticks". Similarly, Wilson's conversations with an employee from the UK's Foreign, Commonwealth and Development Office reveal that online targeting of citizens identified as being at risk of turning to religious extremism is often done because it can be used to show that something is being done, despite evidence that such campaigns have no positive effects or even have negative effects \cite{wilson}.

Thus, an independent, interdisciplinary team of experts whose purpose would be to ensure that the government is fulfilling Züger and Asghari's requirements and educate the public officials wishing to use the infrastructure of online targeting for public policy goals is needed. Such an institution would also reduce the unwanted shift in the power balance to the government's favour since the institution would serve to ensure that the government is not abusing its power. This would increase the public's trust in the system and make them less likely to fear judgement or retribution from the government for what they do in the privacy of their online spaces, thereby making them less likely to self-censor their behaviour. Therefore, an institution imagined in this way would contribute to solving both the Equality and the Privacy Problems. 

In the end, as we have argued, it seems possible to reconcile governmental use of online targeting with democratic principles and values. To do so, the government must satisfy Züger and Asghari's five requirements, while being overseen by an independent organisation yet to be established.

\section{Conclusion}

This paper addresses the socio-political implications of a previously overlooked governmental practice: the use of online targeted advertising for public policy objectives. Our discussion is anchored on a particular case involving the UK government's use of this practice, as explored in the study by Collier et al. \cite{Collier2022}. We argued that this practice, as characterized in this paper, is strikingly undemocratic and raises three major anti-democratic concerns: the Transparency Problem, the Privacy Problem, and the Equality Problem. To reconcile this practice with democratic principles, we sketch the outline of a solution: that the governmental use of online targeting should adhere to an AI governance framework, such as the one developed by Züger and Asghari \cite{zuger2022}, and be monitored by an independent organization comprising interdisciplinary experts.

In order to provide a comprehensive understanding of the topic, it is important to acknowledge the limitations of this paper. Firstly, our analysis is based on Lever's formulation of democratic principles \cite{lever2006}; future work should consider and test other desirable socio-political principles in relation to the use of online targeting by governments. Secondly, we focused exclusively on a single AI governance framework proposed by Züger and Asghari \cite{zuger2022}; future research should explore alternative frameworks and assess their applicability to this context. Lastly, our investigation primarily centers on the practices of the UK government; conducting further studies to examine the use of online targeting in other democratic countries would be beneficial to better understand the generalizability of our findings and recommendations.

While this paper sheds light on the issue of governmental use of online targeting, numerous questions still remain unanswered. For example, is it appropriate for governments to employ these technologies for national security purposes when transparency may not be feasible? Should they also utilize it to counter misinformation and disinformation within online communities? Our objective in writing this paper is to contribute to the ongoing discourse and provide preliminary insights into these complex questions.

\begin{acks}
We would like to thank Ben Collier and James Stewart for their comments on an earlier draft of this paper. We are also thankful to the three anonymous reviewers for their helpful suggestions and feedback.
\end{acks}

\bibliographystyle{ACM-Reference-Format}
\bibliography{sample-base}
\end{document}